\RequirePackage{amsmath}
\documentclass[epj]{svjour}
\usepackage[utf8x]{inputenc}
\usepackage{amsmath}
\usepackage{amsfonts}
\usepackage{amssymb}
\usepackage{graphicx}
\usepackage{mathtools}
\usepackage{subfig}
\usepackage{mathrsfs}

\begin{document}
\title{Anisotropic turbulence in relativistic plasmas.}

\author{Abhisek Saha \and Soma Sanyal}

\institute{School of Physics, University of Hyderabad, Gachibowli, Hyderabad, India 500046}

\mail {somasanyal@uohyd.ac.in}

\abstract{
Signs of turbulence have been observed at the relativistic heavy-ion collision at high collision energies. We
study the signatures of turbulence in this system and find that there are significant departures from isotropic turbulence in the initial and the pre-equilibrium stages of the collision. As the anisotropic fluctuations are subleading to the isotropic fluctuations, the Kolmogorov spectrum can usually be obtained even for the initial stages. However, the energy spectrum and the temperature fluctuations indicate deviations from isotropic turbulence.  Since a strong momentum anisotropy exists between the transverse and the longitudinal plane, we study the energy density spectrum in these two planes by slicing the sphere into different planes. The geometrical anisotropy is reflected in the anisotropic turbulence generated in the rotating plasma and we find that the scaling exponent is different in the two planes.  We also obtain the temperature spectrum in the pre-equilibrium stages. The spectrum deviates from the Gaussian spectra expected for an isotropic turbulence. All these seem to indicate that the large scale momentum anisotropy persists in the smaller length scales for the relativistic heavy-ion collisions. }

\maketitle


\section{Introduction}

The experiments of heavy ion collisions at the Relativistic Heavy Ion Collider (RHIC) in Brookhaven National Laboratory (BNL) and the Large Hadron Collider (LHC) in CERN have successfully produced the Quark-Gluon plasma (QGP)\cite{expt1,expt2,expt3,expt4,expt5}. The QGP is a state of strongly interacting particles (the quarks and the gluons) which is expected to be formed at an extreme condition of temperature and pressure after the quark-hadron phase transition. Subsequently, the system evolves and cools to form the color singlet hadrons which are stable at lower temperatures. The collective behaviour of these produced particles is a sign of the fluid dynamic properties of the system. Based on the hydrodynamic equations, there are models, which can describe the results of these experiments very efficiently. We get similarities in the distribution of elliptic flows at LHC and RHIC which is consistent with the predictions of the viscous hydrodynamic models \cite{viscous1,viscous2,viscous3,viscous4,viscous5,viscous6,viscous7}. In fact, the ideal fluid dynamics can also describe the experimental results exceptionally well \cite{fluid1,fluid2,fluid3,fluid4}.  The fluid produced is considered to be the most perfect fluid as the shear viscosity to entropy density ratio $\eta/s$ is very low. The RHIC data from the top energies favored the value of $\eta/s \approx \frac{1}{8\pi}$ \cite{etas1,etas2}. While the hydrodynamic models are required to follow the equilibrium conditions, the initial stage is far from equilibrium in all respects. Kinetic theory has been used to describe the initial stage as well as the pre-equilibrium stages of the system. The collective flows can also be described by the kinetic theory models and transport models. In some of the transport models, the final stage scatterings are also included which gives a more promising match to the results \cite{kin1,kin2}. The AMPT is one such model based on the kinetic theory which is efficient in producing the rapidity and the transverse momentum spectra that has been experimentally observed. There are also hybrid models that uses the initial stage pre-equilibrium evolution from the AMPT (or any transport model) and takes it as the initial condition at an instant and feeds it to the hydrodynamic models for the near-equilibrium evolution \cite{Bhalerao} of the system. In this work, we use the AMPT model to study the anisotropies in the fluid turbulence both in the initial stage and in the pre-equilibrium stage of the heavy ion collision. By the initial stage here we mean the state of the system after the collision and before the parton scattering. By pre-equilibrium stage we mean the state of the system after the parton scattering but before we can call it an equilibrium state.    

In the initial stages after the heavy ion collision, the system possesses a lot of fluctuations. There are energy density fluctuations, number density fluctuations, temperature fluctuations etc. at different length scales\cite{numberfluc}. These can create disturbances in the flow. We are interested to study the spectra of the initial fluctuations by studying the temperature and velocity distribution in the collision region. While the velocity fluctuation can be studied for the initial stage, we study the temperature fluctuations only in the pre-equilibrium stage where the temperature can be defined using the energy density and the concept of local thermal equilibrium. Since we are using a grid based simulation, it is assumed that in the pre-equilibrium stage local thermal equilibrium exists at the lengthscales given by the grid size. 

There are various reasons for the generation of these fluctuations. In reference \cite{weible1,weible2,weible3}, the authors have shown how the chromo-Weibel instability can be formed in the QGP in the presence of a momentum anisotropy. The Weibel instability can be seen in the homogeneous or nearly homogeneous plasma due to the presence of the momentum anisotropy. This instability can also be generated because of the electromagnetic fluctuation of two opposite moving beams. In ref \cite{tur_chromo1,tur_chromo2}, the generation of turbulent color fields in relativistic plasma is discussed. The onset of turbulence and it's dependence on fluctuations have also been discussed in ref. \cite{floerchinger}. Here the authors have related the full relativistic dynamics of fluctuations to a set of rescaled coordinates for a non-relativistic Navier Stokes equation. This has allowed the authors to use the concepts of non-relativistic turbulence in characterizing the flow in the relativistic heavy ion case. The authors point out, that the Kolmogorov spectrum may not be expected for the correlation functions in the relativistic heavy ion case, but the onset of turbulence can result in a power law structure similar to that obtained for the non-relativistic cases.  

Another interesting point of study is the transition to the equilibrium state. As the system evolves, it tries to reach a state of thermal equilibrium. Because of the geometry of the collision, a large amount of angular momentum is generated in the system. This angular momentum can lead to the generation of vortices in the initial stage plasma\cite{vorticity1,abhi,vorticity2,csernai}. These vortices carry a large portion of the incoming nuclei energy. Thus it can form a turbulent flow. Turbulent flow  consists of many interacting swirls of fluid known as eddies or vortices. These eddies can be of different sizes. Their length scales can be as low as the size of the nuclei and sometimes they can be as large as meters. Turbulence is studied as fluctuations in the velocity field in the laminar flow. The field of the instabilities can cause rapid isotropization and thus can quickly achieve an equilibrium state. This also means that the turbulent flow is diffusive i.e. the energy as well as the momentum spreads out in this fluid. It is also dissipative which means that with time, it dies away gradually. In presence of an active source of energy in momentum space, we get driven turbulence which is stationary. But in heavy ion collisions, we do not have a constant source of energy. Thus, we get a freely propagating energy cascade which is described as a free turbulence. 

In a laminar flow, the temperature at a fixed point is constant for a steady state. However, in the case of turbulent flow, the heat transfer as well as temperature at a point is a function of time. Heat flow is similar to momentum transfer in a turbulent flow where velocity is a function of time. So just like velocity fluctuations, we have fluctuations in temperature in a turbulent flow. In this case, there will be an additional momentum and heat transfer in presence of these fluctuating components. 

Turbulence has been studied in the case of the scalar field theories \cite{sft1,sft2,sft3,sft4} and also in Quantum Chromo Dynamics(QCD) \cite{qcd1,qcd2,qcd3} in recent times. In 1883, Kolmogorov hypothesized that the amount of energy in a turbulent flow carried by eddies of diameter $D$, gravitate towards $D^{5/3}$. But this is only valid within a specific range of length scales known as the inertial subrange. Thus in this range, Kolmogorov spectra will have a power-law nature where the kinetic energy is given by, 
\begin{equation}
	E(k)\approx k^{\nu}\approx k^{-5/3}
	\label{eq:eknu}
\end{equation}
  The Color Glass Condensate (CGC) lattice simulation in ref \cite{cgc1,cgc2} has shown that the transfer of energy from low to high momenta follows a power-law behaviour with the exponent $\nu=-5/3$. In general, the exponent $\nu$ has a wide range of validity. In the case of renormalizable classically scale invariant interactions in QCD, $\nu$ has been found out to be $-5/3$ and $-4/3$ for energy and particle cascade respectively \cite{carrington}.

Generally, the turbulence associated with the relativistic heavy ion collision is assumed to be an isotropic turbulence \cite{isotropicturbulence}. There have been proposals where the turbulence proceeds not through an energy cascade but by an entropy one \cite{calzetta}. These are also studied under the assumption of homogeneity and isotropy. The author has shown that a fully relativistic turbulence has a richer dynamics compared to the non-relativistic case. They have studied the tensor driven turbulent flow for a relativistic, isotropic and homogeneous fluid. However, the patterns they have mentioned are difficult to reproduce in the actual situation. It is also well known that the system has a momentum anisotropy and is rotating. In this work, we are interested to study the anisotropies generated in the turbulence. Since there is a strong momentum anisotropy in the rotating system, we, therefore, use the same planes to study the anisotropy in the turbulence spectra. We find different scaling exponents in the transverse ($x-y$) and the longitudinal ($x-z$)  plane. In the $x-y$ plane, the exponent is close to the viscous convective range while in the $x-z$ plane, it is closer to the Kolmogorov convective range. The Kolmogorov type exact scaling relation usually holds for the non-relativistic case. Though currently many studies have indicated that the traditional interpretation of the energy cascade in the Kolmogorov case may be misleading, it has been shown that exact scaling results can be obtained for even relativistic turbulence, which reduces to the Kolmogorov relation for lower velocities \cite{fouxon}. A recent work also established the existence of a relativistic energy cascade in the traditional sense \cite{eyink}. This indicates that the relativistic turbulence can also be studied using the energy cascade model. In recent times, it has become important to study the role of vorticity and other phenomena in relation to the turbulence generated in the initial stages of the heavy ion collision \cite{abhi,csernai}. The AMPT model, gives the distribution of relativistic velocities of the particles formed after the collision. We have already used it to study the vorticity patterns in the initial stages of the plasma\cite{Abhisek_vorticity}. The velocity correlations can be obtained from the output velocities given by the simulation. In this work, we have used the relativistic output from the AMPT simulations to obtain the energy spectrum in the initial stages through velocity correlation. As it is known that the initial geometry of the collision is anisotropic, we have analyzed the anisotropic turbulence using the energy cascade, however, the velocities involved are relativistic velocities. The tensor degrees of freedom often show different features for relativistic and non-relativistic velocities but in this work we do not study these higher degrees of freedom.

Anisotropic turbulence can also be studied by studying the temperature fluctuations of the turbulent fluid. The temperature spectrum of a turbulent fluid has been discussed in the literature \cite{corrosin}. For an isotropic turbulence, the spectrum of the temperature fluctuations is found to be a Gaussian. Deviations from the Gaussian spectrum, indicate the presence of anisotropies in the turbulent system. However, the temperature fluctuation spectrum has not been studied in any of the turbulence studies in relativistic heavy ion collisions. This spectrum can tell us about the thermal lengthscale of the system. The smallest lengthscale associated with this spectrum is also related to the size of the smallest eddies. This is studied in the pre-equilibrium stages of the heavy ion collision. The reason for this is that  we calculate the temperature from the energy density. We are assuming local thermal equilibrium  at various points in the system.  Unlike the case of velocity fluctuations here we are limited to the whole three dimensional modeling of the spectrum so we consider the spectrum as a whole for all the temperature fluctuations. In usual cases, the 3-dimensional spectrum of temperature fluctuations is a Gaussian for an isotropic turbulence. So we try to obtain a similar Gaussian spectrum for the case of the relativistic heavy ion collision. We find that though the spectrum can be approximately fitted with a Gaussian, there are several interesting deviations. We find that the Gaussian peak of the spectrum shifts from higher length scales to shorter length scales with time. At later times the spectrum is a better fit with the q-Gaussian distribution. We use the spectrum to find the shortest length scale of the eddies in the turbulent flow. We find that the spectrum of temperature fluctuations remains more or less the same independent of the collision energy.

In this paper, we have analyzed the power spectrum of turbulent flow in relativistic heavy ion collisions. To obtain the power spectrum of turbulent flow, we use the initial stage parton distribution given by the AMPT model. The velocity spectrum is obtained by separating the fluctuating component of the velocity. The spectra for the turbulent flow in the transverse and the longitudinal plane are studied. We have discussed the range of wave number and eddy sizes at which the inertial range and the viscous range dominates in each of the planes for different initial conditions. For the pre-equilibrium stage, we have looked at the temperature spectrum. The temperature spectrum has several non-Gaussian features. We have analyzed how the power spectra and the temperature spectra both indicate the presence of an anisotropic turbulence in the rotating system in the initial and the pre-equilibrium stages. 

In section 2, we describe the AMPT model. Section 3 discusses the formulation of the turbulence energy spectra. Here we have also shown the range of eddies we can get and the range of wave numbers used in order to consider all the dissipation as well as inertial eddy sizes. In section 4, we show the results obtained for the numerical simulations. We have shown both the longitudinal as well as transverse energy spectra at different initial conditions. In section 5, we describe the power spectrum of temperature fluctuations that we obtain for the turbulent system. We summarize the results and conclude this paper in section 6.

\section{The AMPT model}
The AMPT is a publicly available simulation of the relativistic heavy ion collision \cite{ampt} which has been widely used as an event generator. We use this model to get the initial state pre-equilibrium dynamics of the collision system. The initial state is generated using A Heavy Ion Jet Interaction Generator(HIJING)\cite{hijing}. In the string melting version of this model, the hadrons coming from the HIJING generator are converted into valence quarks and their collision dynamics are modeled by Zhang’s Parton Cascade (ZPC) model\cite{zpc}. The hadronization is described by a quark coalescence model. It gives mesons by converting two nearest quark and anti-quark pairs and baryons (anti-baryons) by converting three nearest quarks (anti-quarks). The successive hadrons collide until freeze-out is described by, A Relativistic Transport (ART) \cite{ART} model. We have used the AMPT model to obtain the velocity correlations in the initial stages. Later on when we do the temperature fluctuations we have looked at the partons after scattering to obtain the energy in the pre-equilibrium stages.  We will give all these details in the relevant sections. First we discuss the kinetic energy spectrum from the velocity correlations in the next section, later on we will give the details of the equations pertaining to the temperature spectrum in section 5.

\section{The turbulence spectra}

The geometry of the heavy ion collision ensures that the velocity flow becomes non - uniform in the rotating fluid. This gives rise to turbulence in velocity field in the initial stage. The turbulent flow consists of eddies of various sizes.  
To get the turbulent component of the velocity field, we have to separate out the laminar component from the actual velocity. If we divide the actual velocity into two parts, the laminar flow and the fluctuating component, we can get the turbulent component as, 
\begin{equation}
	\vec{u}(\vec{x})=\vec{U}(\vec{x})+\vec{u}^{'}(\vec{x})
\end{equation}

Here $\vec{U}=\left<\vec{u}\right>$ is the laminar component and $\vec{u}^{'}=\vec{u}-\left<\vec{u}\right>$ is the turbulent component. The vector $\vec{x}$ gives the position vector for the point at which the velocity is considered. In our case we use a discretized grid structure to determine the average velocity and hence in those cases, the position vector denotes the position vector of the cell whose average velocity we are considering. We will discuss this in more detail when we discuss the velocity correlations later in this section. Throughout the paper, any primed quantity will denote the turbulent component of the particular field. 
As mentioned in the previous section,  we get the individual particle momenta from the AMPT model. We use this to get the laminar as well as the turbulent component of the velocity. In the case of turbulence study, we do a statistical average of the velocity. This gives a more deterministic solution than the study of velocity through the Navier-Stokes equations. There are two types of statistical averages to obtain average velocity.  For the first one, the space average is obtained where velocities are considered at a fixed time and the average is taken over the whole volume $V$ occupied by the system.
\begin{equation}
	\left<\vec{u}\right>=\lim_{\Delta x\rightarrow 0}\int_{V} \frac{\vec{u} d^{3}x}{V}
\end{equation}
And the other one is the time average, where the position in space is kept fixed and the averaging is done over time. 
\begin{equation}
	\left<\vec{u}\right>=\lim_{\Delta T\rightarrow \infty}\int_{0}^{T} \frac{\vec{u} dt}{T}
\end{equation}

In the current work, we are interested to see whether the anisotropy in the initial geometry of the relativistic heavy ion collisions is reflected in the  turbulence spectrum of the initial stages of the collision. For this, we use the velocity correlation tensor for the turbulent velocity given by, 
\begin{equation}
R_{i j} (\vec{r}) = \left\langle u_i'(\vec{x}) u_j'(\vec{x} + \vec{r}) \right\rangle
\label{eqn:velcorre}
\end{equation}
Here $u_i'$ denotes the fluctuating part of the velocity. We are using a grid based method and the $\left\langle \right\rangle $ denotes the average over space (i.e $\vec{x}$). This gives the  correlation between the fluctuating velocities at the two points denoted by $\vec{x}$ and $\vec{x} + \vec{r}$.  
The $R_{ij}$ is related to the energy spectrum tensor $E_{ij}(\vec{K})$ by, 
\begin{equation}
E_{i j} (\vec{K}) = \frac{1}{(2 \pi)^3} \int \int \int e^{-i\vec{K}.\vec{r}} R_{i j} (\vec{r}) d(\vec{r}).
\label{eqn:engspectrum}
\end{equation}
The AMPT code is used to obtain the positions and velocities of the particles. After that we have  calculated the Fourier transformation of the space coordinates to the corresponding Fourier wave vector space for the heavy ion collision geometry.

The velocity correlation can be expressed as a second order tensor. This is given in equation \ref{eqn:velcorre}. If isotropic turbulence is assumed then, the Fourier transform to the wave vector $\vec{K}$ space involves all the three dimensions and is finally expressed as $E(k)$ where $k$ is the magnitude of $\vec{K}$. $E(k)$ is therefore the three dimensional kinetic energy spectra. It is given by, 
\begin{equation}
E(k) = \frac{1}{(2 \pi)^3} \int \int \int e^{-i\vec{K}.\vec{r}} R_{i j} dx dy dz.
\label{eqn:scalarspectrum}
\end{equation}
where $dx dy dz$ is the volume element over which the integration is carried out.  

For a lower dimensional energy spectra that is only a profile of the complete spectrum, the usual method is to choose a  preferred axis and to determine the fluctuating velocity components parallel to this axis. This gives us the longitudinal spectra. For the transverse spectra, the velocity correlation tensor is obtained for the velocity fluctuations orthogonal to this axis.  It can be shown that the longitudinal spectrum and the transverse spectrum in an isotropic turbulence have the same coefficient in both the planes. For relativistic heavy ion collisions we choose the preferred axis as the $z$ axis. This means that to obtain the longitudinal spectrum we need to find the velocity correlation in the plane parallel to the $z$ axis (i.e., the x-z plane) and for the transverse spectra we need to find the velocity correlation  perpendicular to the  $z$ axis. 
The only challenge here is that since the particles are colliding with relativistic velocities along the  $z$ axis, we need to take care of the Lorentz boost effect when we calculate  the velocity correlation in the longitudinal plane. We will discuss this in more detail in the next section where we discuss and give results for the two spectra separately.

The energy spectrum basically shows how the kinetic energy is distributed over the different eddy sizes. The eddy sizes are determined by the length scales in the system so before we proceed to discuss the turbulent spectra in detail, we briefly discuss the length scales involved in this particular system.

 The structures of the rotating elements in a turbulent system can be of different sizes. Thus we have different length scales in our problem. The length scale of a system is the distance over which the characteristic gradients of different variables are present. The largest eddy formed in the system can be of the largest length scale of the system. These large eddies extract kinetic energy from the mean flow and use it to develop angular momentum. In relativistic  heavy ion collisions, most of the energy gets converted into angular momentum and is eventually dissipated through the smaller eddies. This is known as the energy cascade. This energy cascade can be expressed in terms of the Reynolds number. The Reynolds number is defined as the ratio of inertial force to the viscous force\cite{book};
\begin{equation}
	Re=\frac{F_{i}}{F_{v}}=\frac{\rho ul}{\mu_{d}}
	\label{Re}
\end{equation}
Here $F_{i}=\rho l^{3}\frac{u^{2}}{l}$ is the inertial force and $F_{v}=\mu_{d}\frac{u}{l}l^{2}$ is the viscous force. $\rho$, $l$ are the density and the length scale and $\mu_{d}$ is the dynamic viscosity of the fluid. Thus, the onset of turbulence depends on the density and viscosity of the fluid, size of the medium and velocity of the fluid. In case of  a large Reynolds number, the fluid viscosity is less dominant over the fluid inertia and  we get larger eddies. This is the regime of the Kolmogorov spectra \cite{kolmo1}. Here the vortex interaction describes the mediation of scale invariant flow of spectral energy. The energy gets transferred from the mean flow to the large eddies and then it dissipates through the smaller eddies \cite{wavetur1,wavetur2,wavetur3}. In relativistic heavy ion collisions, we typically have large Reynolds numbers, hence we expect the energy spectrum to be of the Kolmogorov type. The largest length-scale in the system is dictated by the system size.  
In the simulations, we have taken a cell size of $0.3$ fm and $48$ cells in each direction. So, our system size is $l=14.4$ fm in each direction. Also the diameter of the $Au$ nuclei is around $12$ fm.  Hence, this will be the largest eddy size in the system. Thus, this corresponds to the minimal wave number $k_{min}$.
\begin{eqnarray}
	k_{min}=\frac{2\pi}{l}=0.524 \;fm^{-1}
	\label{eq:kmin}
\end{eqnarray} 
In case of minimum bias $Au-Au$ events, impact parameters are taken in the range of $0-15$ fm. In that case, $k_{min}\approxeq 0.42\;fm^{-1}$.

 The Kolmogorov length scale is defined as the length scale of the smallest eddy. This can be found out by making the Reynolds number very small in Eq.\ref{Re} \cite{book}. It is given by,
\begin{equation}
	\zeta=\left(\frac{\mu_{k}^{3}}{\epsilon_{d}}\right)^{1/4}
	\label{eq:klen}
\end{equation}
Here $\mu_{k}$ is the kinematic viscosity and $\epsilon_{d}$ is the energy dissipation rate.
The kinematic viscosity is related to the dynamic viscosity by the relation $\mu_k = \frac{\mu_d}{\rho}$ If we put the dimension of $\mu_k$ and $\epsilon_d$ in Eq. \ref{eq:klen}, we will get the dimension as the length scale\cite{book}. The smallest length scale is related to the Reynolds number by,
\begin{equation}
	\zeta = l Re^{-3/4}
	\label{eq:eta}
\end{equation} For a QGP system, the kinematic viscosity obtained in ref. \cite{viscosity} is,
\begin{equation}
	\mu_{k}\approx10^{-7}\frac{m^{2}}{s}\approx1.69 \;GeV^{-1}
\end{equation}

Now, in order to create a QGP state in the heavy-ion collision, the energy density must exceed the nucleonic density. If we take this energy density bound around $2\;GeV/fm^{3}$ then from Eq. \ref{eq:klen}, the Kolmogorov length scale is $1.24$ fm. As this is the lowest eddy scale, we can get the corresponding value of the wave number similar to equation \ref{eq:kmin}. So, in this case, $k_{max}\approx 5$. In ref. \cite{brett}, it is found that the Reynolds number $Re\geqq 8.52$ for the system created in RHIC energies. In that case, $\zeta\approx 2.4 \;fm$ which is obtained using Eq. \ref{eq:eta} and the minimum value of $k_{max}$ is $3$. The minimum length scale can vary depending on the choice of planes. Also, for lower Reynolds numbers, smaller eddies can be formed. In the simulations, we use the range of the energy spectrum from $k = 0.5$ to $k = 20$ to include all possible length scales of the $Au-Au$ collisions at RHIC energies.

\section{Results and Discussions}
\subsection{Longitudinal plane spectra}

We first discuss the case of the longitudinal plane. The output of the AMPT gives us the velocities along with the positions of the particles. This plane includes the $z$ axis. 
This means that we are considering the longitudinal velocity correlation between two points on the $x-z$ plane which are at a distance $\vec{d}$ from each other, the correlation will be given by 
\begin{equation}
R_{ij} = <u'_i (\vec{r},t), u_j'(\vec{r} + \vec{d},t)>
\label{eq:longicorrel}
\end{equation}
Here the $\vec{r}$ is the position of grid cell with turbulent velocity $u_i'$ and $\vec{r} + \vec{d}$ is the position of the grid cell with velocity $u_j'$ 
We consider equal time correlators and drop the "t" in future equations. The only problem that arises here is due to the Lorentz boost along the $z$ axis.
So the $\vec{d}$ is boosted along the $z$ axis. This means that the correlation function has to be replaced to take care of the Lorentz boost. We found this worked out in a recent paper \cite{anbasar} and have used it to calculate the energy spectrum in our case. In our case, we only need the equal time correlator for the relativistically boosted velocities. So we need,  
 \begin{equation}
R_{ij} =  \Lambda(d/2) \Lambda(-d/2)  <u'_i ({r - d/2 }), u_j'(r + d/2)>
\label{eq:boosted}
\end{equation}
where now the vector $\vec{d}$ is the line joining the two positions given in Eq.\ref{eq:longicorrel} and the correlator is boosted to the local reference frame of the midpoint between the two points whose correlation we are interested in.
So basically we will have results independent of our choice of reference frames. 
This is difficult to implement in the grid structure that we have. We implement it by assuming that it is an infinitesimal boost. This is also discussed in detail in ref.\cite{anbasar}. We obtain the $\Lambda(d/2)$ matrices for an infinitesimal boost and proceed to calculate $R_{ij}$ for each pair of velocities in the $x-z$ plane. Once we obtain $R_{ij}$ we proceed to calculate the 2 - dimensional scalar kinetic energy spectrum $E(k)$ defined in Eq.\ref{eqn:scalarspectrum}. The integration is carried out between the minimum and maximum lengthscales of only two dimensions. The boost matrices take care of the contraction of  the lengthscale in the direction of motion. So instead of integrating over the position $r$. the integration is done for the position $r+\frac{d}{2}$. We have also considered the real part of the exponential for our calculations.

As mentioned before, generally for studying a profile of the complete three dimensional spectrum, there is a chosen axis and the planes are chosen parallel and perpendicular to the chosen axis. So the three dimensional energy spectrum $E(k)$ will be replaced by the energy spectrum in the parallel plane which we denote by $E_{long}(k_z)$ and by the transverse energy spectrum in the perpendicular plane which we denote by $E_{tr}(k_{tr})$. It should be noted as mentioned before only the magnitude of the wavevector is considered in these spectra and the notations $k_z$ and $k_{tr}$ denote the scaler magnitudes of the vectors and do not refer to any components in any direction.  
 
The exponent of $k_z$ is determined after fitting the graphs for $E_{long}(k_z)$ obtained from the simulation for different values of $k_z$. It is denoted by $\nu$. 
\begin{figure}
	\includegraphics[width = \linewidth] {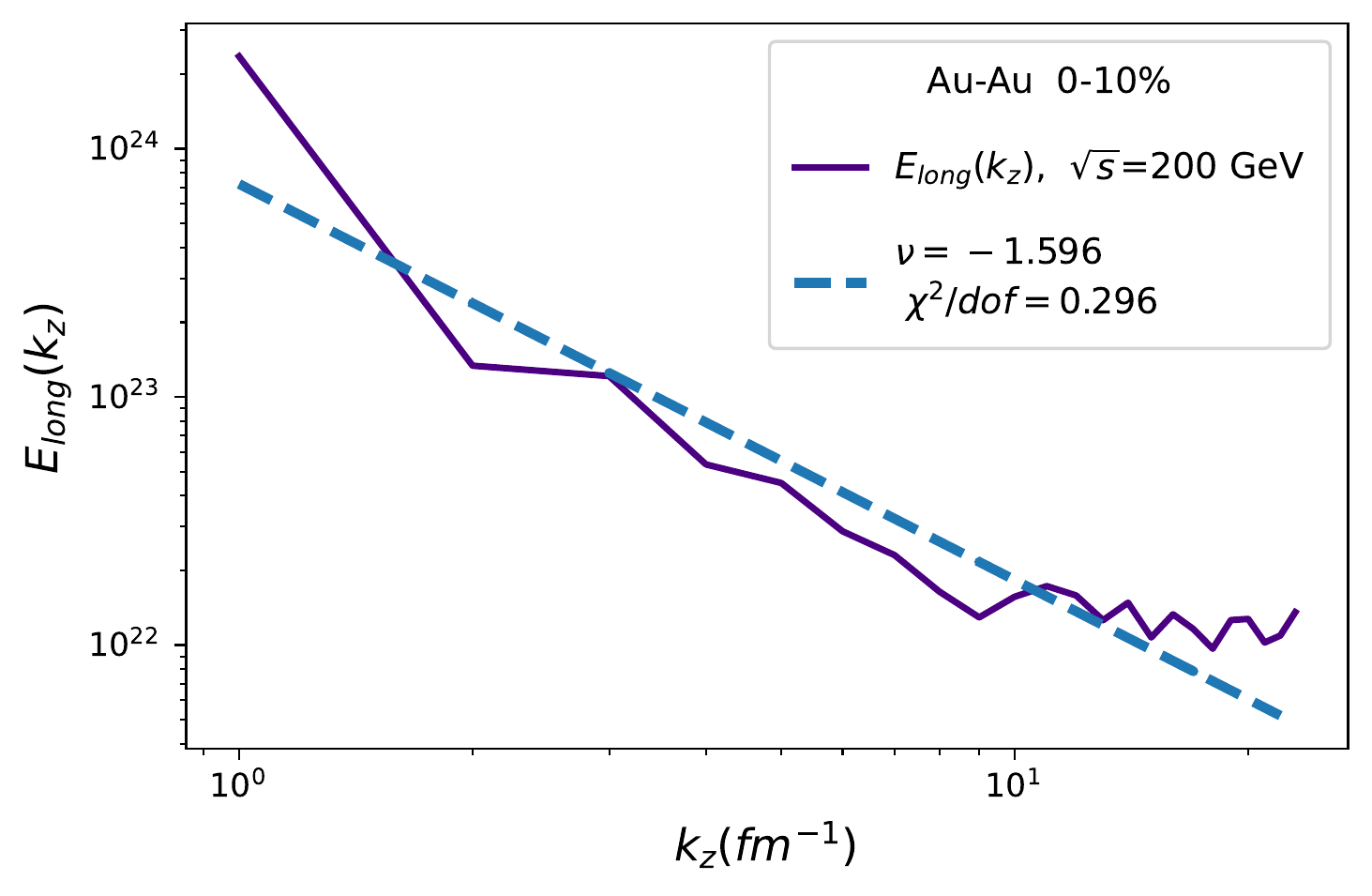}
	\caption{Turbulence velocity spectra at collision energy 200 GeV. The velocities are considered on the longitudinal plane. The range of centrality is $0-10\%$. $\nu$= -1.596  }
	\label{fig:200_10}
\end{figure}
\begin{figure}
	\includegraphics[width = \linewidth]{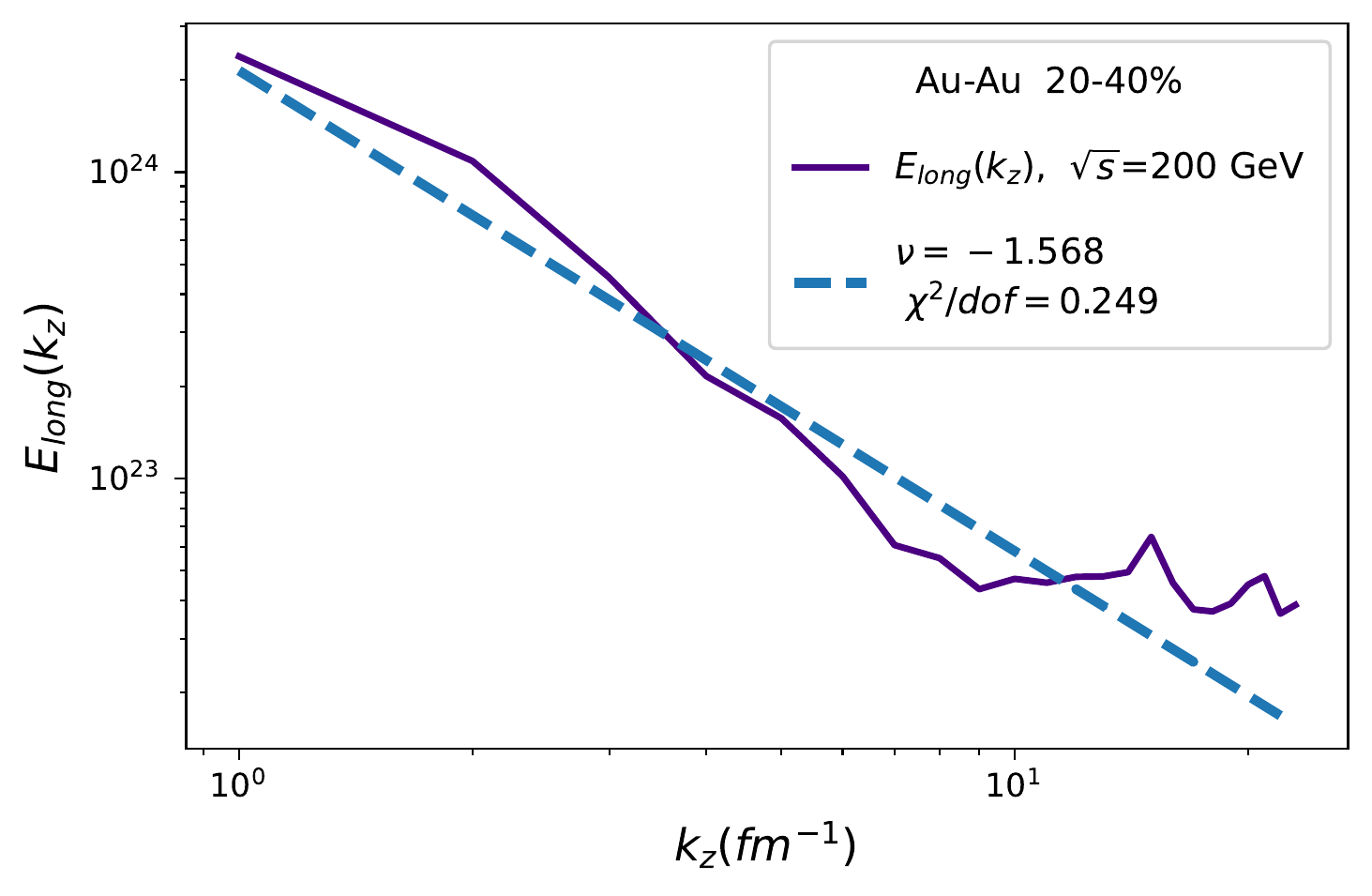}
	\caption{ Turbulence velocity spectra at collision energy 200 GeV. The velocities are considered on the longitudinal plane. The range of centrality is $20-40\%$. $\nu$= -1.568 }
	\label{fig:200_40}
\end{figure}
 \begin{figure}
	\includegraphics[width = \linewidth]{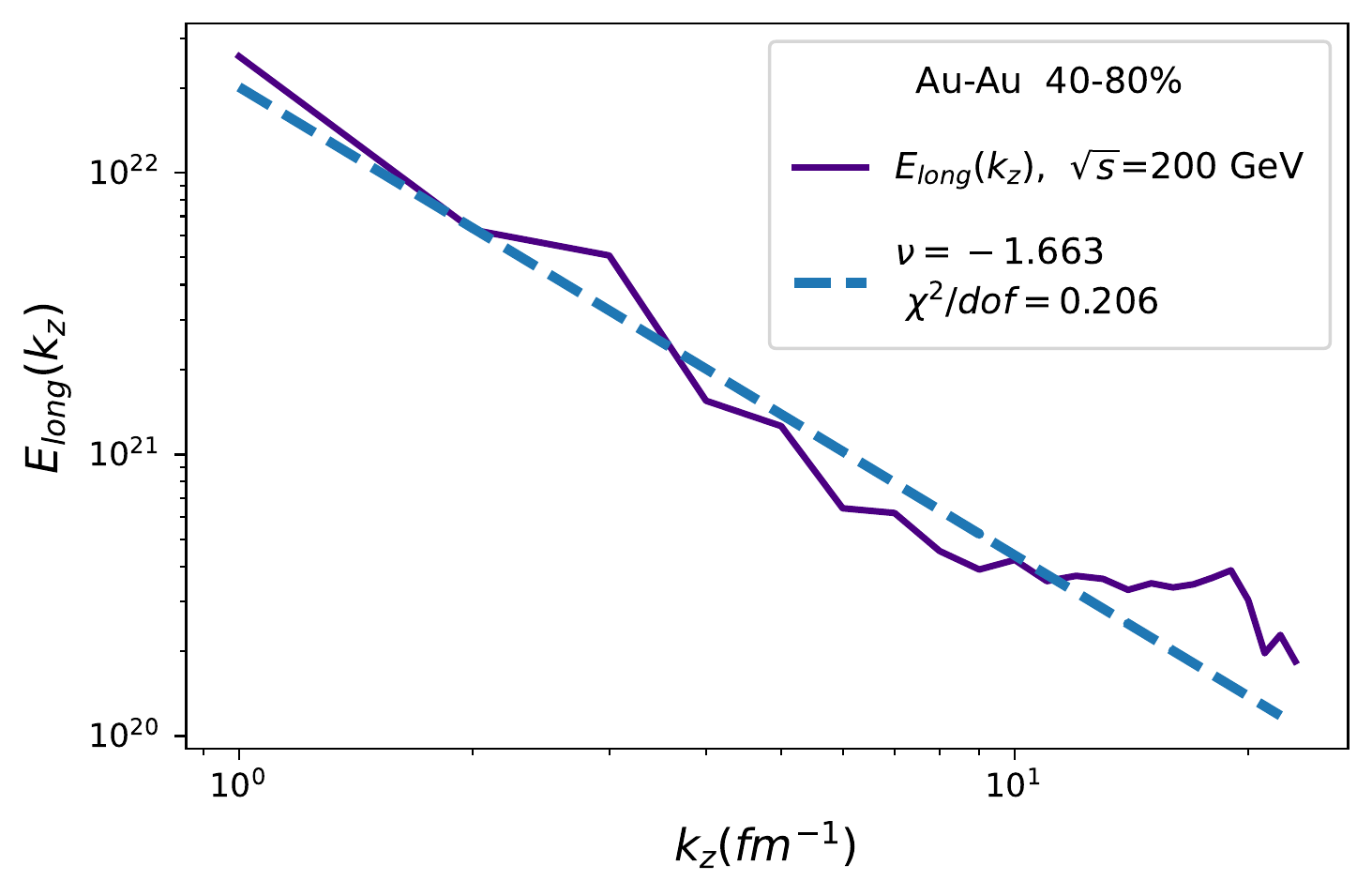}
	\caption{Turbulence velocity spectra at collision energy 200 GeV. The velocities are considered on the longitudinal plane. The range of centrality is $40-80\%$. $\nu$= -1.663    }
	\label{fig:200_80}
\end{figure} 
  Fig. \ref{fig:200_10}, Fig. \ref{fig:200_40} and Fig. \ref{fig:200_80} show the turbulence spectrum of the velocity field at a collision energy($\sqrt{s}$) of $200$ GeV for $0-10\%$ centrality,  $20-40\%$ centrality and  $40-80\%$ centrality. These plots are taken on the longitudinal plane and it is shown in the log-log scale. The dashed line is the best fitted line for the range of $k_z$ considered. From the slope of the fitted straight line, we compute the power $\nu$ (ref Eq. \ref{eq:eknu}). The coefficient of the power spectrum is around $-1.6$ in all the cases. This is approximately equal to $-5/3$. 
This is near to the Kolmogorov limit. So, in this plane, the spectra resembles the Kolmogorov spectrum. Here the inertial force is greater than the dissipative force.

\subsection{Transverse plane spectra}
We now give the results for the spectrum on the transverse plane. The transverse plane is perpendicular to our chosen axis (the $z$ axis). Since the system is boosted only along the z-axis, the components of momentum perpendicular to this axis remains the same. So we do not need to do the boost transformation for the velocities to obtain the velocity correlation tensor. We find the velocity correlation tensor for all the pairs of points and obtain the different values of $R_{i j}$. Using these $R_{i j}$, we obtain the spectrum $E_{tr}(k_{tr})$. 

\begin{figure}
	\includegraphics[width = \linewidth ] {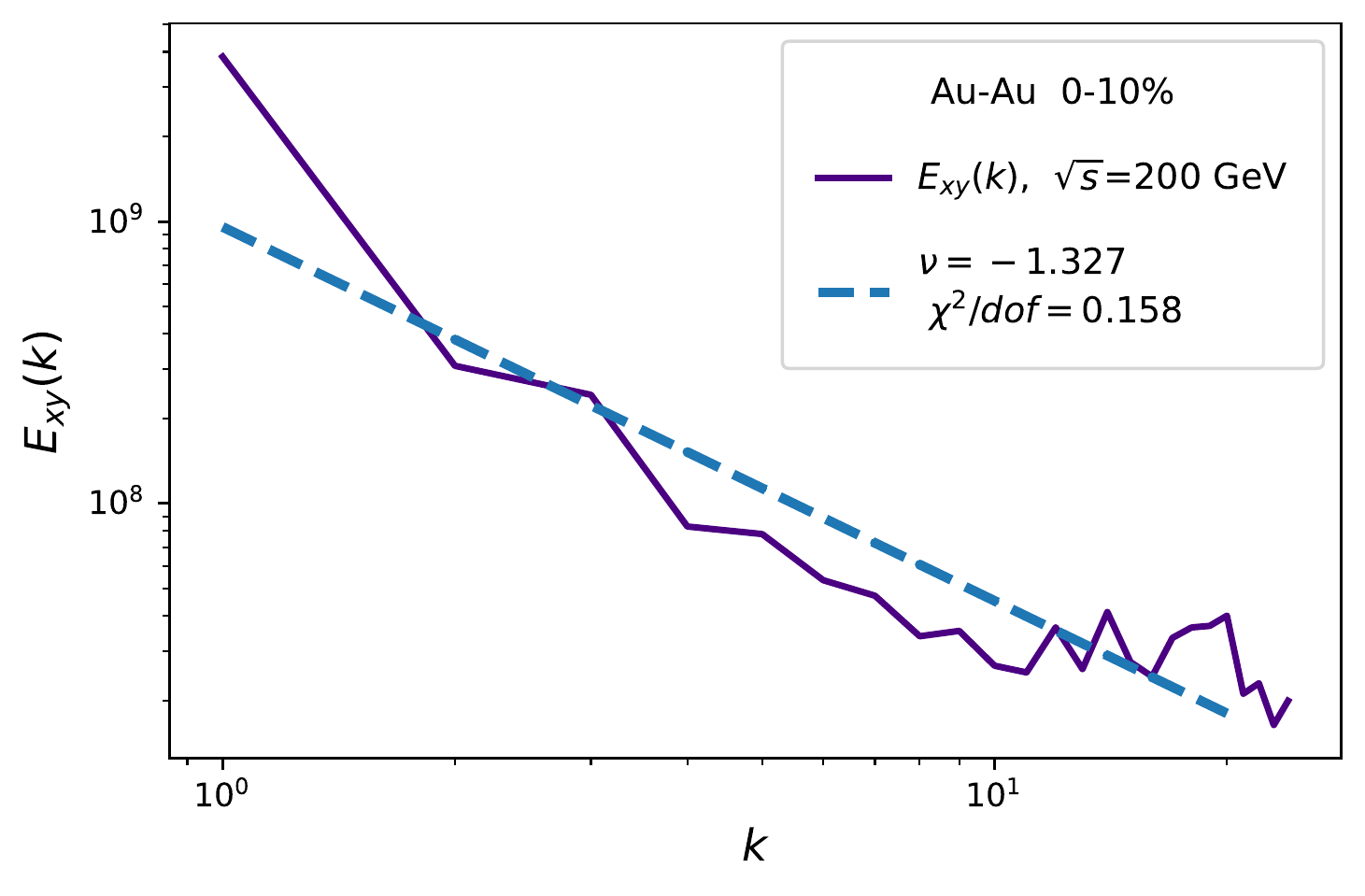}
	\caption{Turbulence velocity spectra at collision energy 200 GeV. The velocities are considered on the transverse plane. The range of centrality is $0-10\%$. $\nu$=-1.264 }
	\label{fig:transverse200_10}
\end{figure}

\begin{figure}
	\includegraphics[width = \linewidth]{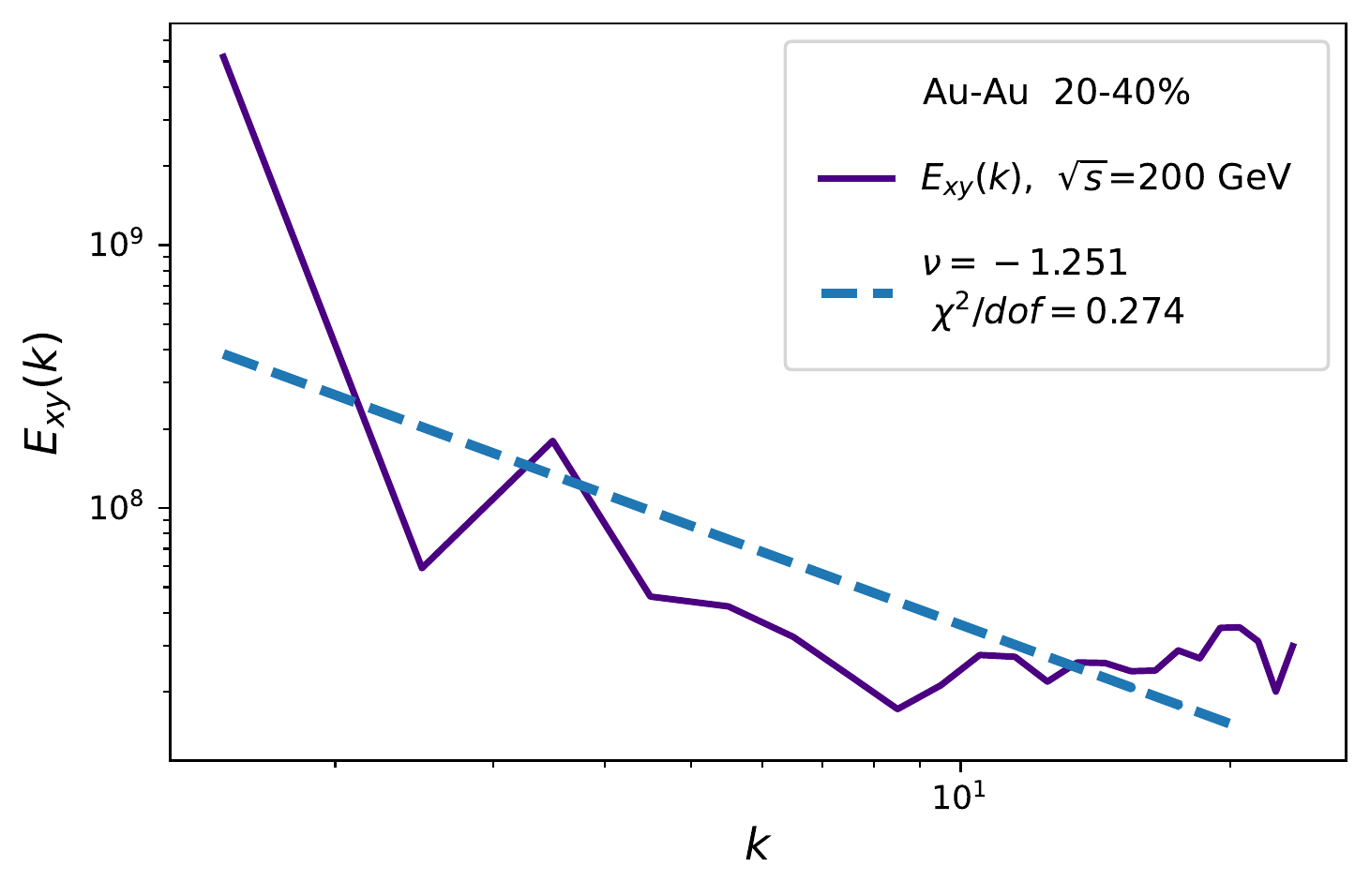}
	\caption{ Turbulence velocity spectra at collision energy 200 GeV. The velocities are considered on the transverse plane. The range of centrality is $20-40\%$. $\nu$=-1.183 }
	\label{fig:transverse200_40}
\end{figure}

\begin{figure}
	\includegraphics[width = \linewidth]{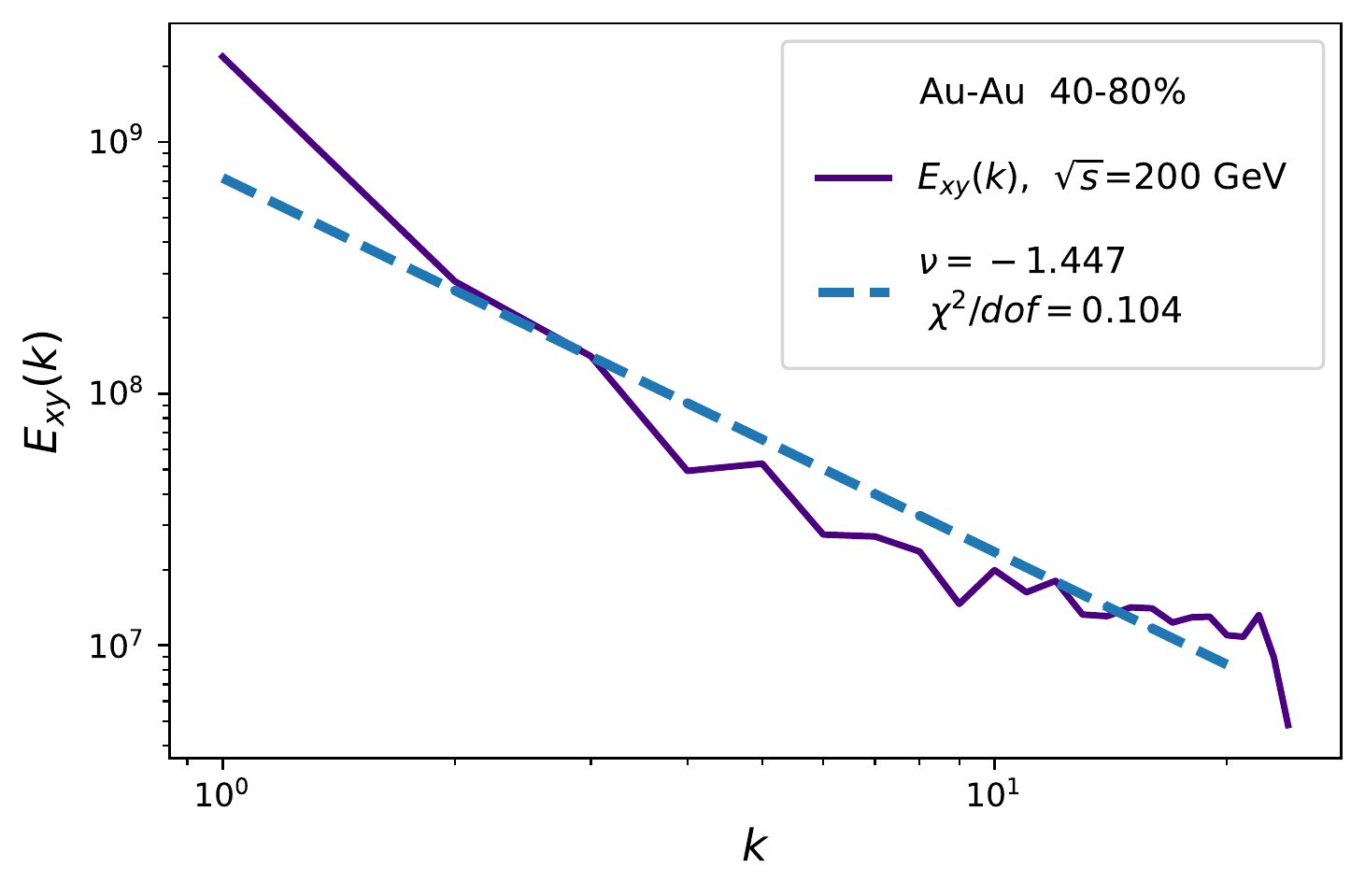}
	\caption{ Turbulence velocity spectra at collision energy 200 GeV. The velocities are considered on the transverse plane. The range of centrality is $40-80\%$. $\nu$= -1.389 }
	\label{fig:transverse200_80}
\end{figure}

Fig. \ref{fig:transverse200_10} and Fig. \ref{fig:transverse200_40} show the turbulence spectrum of the velocity field for $0-10\%$ centrality and $20-40\%$ centrality at the collision energy of 200 GeV. These plots are taken on the transverse plane. The power law exponent $\nu$ is again obtained from the fitted graphs. In Fig. \ref{fig:transverse200_10}, a power law exponent of $-1.26$ is found while in Fig. \ref{fig:transverse200_40}, the exponent obtained is lower in value ($-1.18$). In Fig. \ref{fig:transverse200_80}, the spectrum is given for the $40-80\%$ centrality region and the exponent found is higher than that of $0-10\%$ centrality but is lower than the exponent of the Kolmogorov spectrum.
 Interestingly, the power law exponent does not remain the same in all these cases. A power law exponent of $-4/3$ is obtained when dissipative forces are important. These values appear to be closer to the exponent of $-4/3$ than the Kolmogorov value. We have also measured the exponent for the collision energies at $19.6$ GeV, $62.4$ GeV, $100$ GeV and $130$ GeV. We observe the same nature of the exponents in the different planes. We find that it is always closer to $-5/3$ for the longitudinal plane and closer to $-4/3$ for the transverse plane. This gives rise to the different slopes in the log-log plot. This seems to indicate that our results are independent of the collision energy at this stage. 
 
We have also used a very narrow range of $\eta$ ($\eta$ being the pseudorapidity) in obtaining the energy spectrum in the transverse plane, limiting it to $-0.5 < \eta < 0.5 $.
The longitudinal spectra and the transverse spectra therefore have different coefficients. From all these results, we can infer  that the initial distribution is not spherical. 

Though the difference may not be very significant they will give rise to different Schmidt numbers. The Schmidt number is the ratio of the kinematic viscosity and the inter particle diffusion rate \cite{schmidt}. The transverse spectrum corresponds to a higher Schmidt number than the longitudinal spectra. As is well known the shear viscosity is a measure of the mean free path of a system and the entropy density can be considered a measure of the inter particle distance in heavy-ion collisions. For heavy-ion collision the ratio $\frac{\eta}{s} < 1$. This means that the spectrum of the QGP should resemble a Kolmogorov spectrum. Though this happens in the longitudinal plane, it does not happen in the transverse plane. 

The reason for this is because of the geometry of the collision, the QGP appears "lumpy" in the transverse plane \cite{busza}. After the collision, there is an overlap region shaped like an almond. This overlap region has a very high energy density and large anisotropic pressure gradients in the $x-y $ plane. So the particles are not distributed uniformly on this plane. There is an anisotropy in the particle distribution. 
This means that neither the mean free path nor the inter-particle distance are uniform on this plane. This leads to the different coefficients in the turbulence spectra.  The difference in the coefficient of the power law indicates that the turbulence in the rotating fireball is not isotropic. This anisotropy can be related to the spatial anisotropy of the overlap zone in the two different planes.
The spatial anisotropy ensures anisotropic pressure gradients in the $x-y$ plane \cite{bhalerao}. The difference in the ratio of the viscous diffusion rate to the inter particle diffusion rate in the two planes results in the difference in the coefficient of the power law.    

We have also found that the difference of these coefficients will be reflected in the different values of the Schmidt number for each of the plots in the transverse plane. The higher Schmidt number indicates that momentum diffusion dominates in the transverse plane. It has already been shown \cite{heinz} that the viscous effects which reduce the magnitude of the elliptic flow depend upon the collision centrality. This is reflected in our spectrum as the power law exponent is different in different centralities. There are anisotropic pressure gradients in the $x-y$ planes \cite{bhalerao} which give rise to the difference in the power law exponents obtained in that plane. 

One of the ways of measuring the anisotropy in a rotating turbulent flow in non-relativistic fluids is by defining an anisotropic parameter. The anisotropic parameter is defined by $A = \frac{E_p}{2 E_z}$ where $E_p = E_x + E_y$. The energies are defined in the perpendicular planes of the rotating fluid. In relativistic heavy-ion collisions also, we can define such an anisotropy parameter based on the initial geometry of the system. In this case, the perpendicular planes would be the longitudinal plane and the transverse plane. 
The flow of energy in the system will depend on this anisotropic parameter $A$. We hope to pursue the inclusion of an anisotropic parameter in the development of turbulence in the relativistic heavy-ion collision system in a later work.

 
 \section{Power spectrum of temperature fluctuations}
 
Temperature fluctuations have previously been discussed in heavy-ion collisions in ref. \cite{abhitemp,hotspots,basu} in many different contexts. Once the particle distribution is known, we divide the system into smaller cells. Each of these cells have a reasonable number of particles so that local thermal equilibrium can be assumed in these cells. The temperature of each of these cells can then be obtained using the Gibbs - Boltzmann formula relating the energy density to the temperature. We find that the temperature so obtained has a lot of fluctuations. These temperature fluctuations are expected in heavy-ion collisions \cite{abhitemp}. Now, it is well established that the complete spectrum of temperature fluctuations for an isotropic turbulence is Gaussian in nature \cite{corrosin}. So in our study, we  obtain the power spectrum of the temperature fluctuations in the pre equilibrium stage of the plasma. At that stage, we can assume local thermal equilibrium for the plasma. We find that though, it is possible to fit the temperature spectrum by an approximate Gaussian curve, there are fluctuations present which indicate the presence of anisotropies. This is further enhanced at later times when the spectrum is better fitted with a Poissonian q-Gaussian distribution.

In case of turbulent flow, the shear stress or Reynolds stress can be obtained from the equation of motion. The tangential stress depends on the velocity change perpendicular to the flow direction and the  viscosity of the fluid. 
\begin{equation}
	\Pi =-\rho(\gamma+\epsilon_m)\frac{\partial \bar{v_{x}}}{\partial y}
\end{equation}
The equation comes from the conservation of momentum where $\rho$ is the density, $\gamma$ is the thermal diffusivity coefficient, $\epsilon_m$ is the coefficient of turbulent viscosity and $\bar{v_{x}}$ is the average velocity along the $x$ axis. Similarly, the total heat flow can be obtained from the conservation of energy.
\begin{equation}
	Q =-\rho c_{p}(\alpha+\epsilon_p)\frac{\partial \bar{T}}{\partial y}
\end{equation}
Here $c_p$ is the specific heat at constant pressure,$\alpha$ is the coefficient of thermal conductivity, $\epsilon_p$ is the coefficient of eddy diffusivity for heat and $\bar{T}$ is the average temperature. The heat flow occurs due to the presence of the temperature gradient in the direction perpendicular to the flow direction. 
The above two equations are similar in nature. In both the equation, the first term in the right side is the laminar component and the second term is the turbulent contribution which is composed of two fluctuating components. In the first case, it is the velocity that we have already discussed in the previous sections, for the second case it is the temperature.  We now proceed to study the temperature fluctuations. 

For the case of isotropic turbulence, it is possible to define the power spectrum of temperature fluctuations beginning from the heat transfer equation \cite{corrosin}, 
\begin{equation}
\frac{\partial T }{\partial t} + v_k \frac{\partial T}{\partial x_k}  = \gamma \frac{\partial^2 T}{\partial x_j \partial x_j} 
\end{equation}
Here we have assumed that the temperature at any two points $P$ and $P'$  are different. One can then define $m(r) = <TT'>$ as the temperature correlation between the two given points. The temperature fluctuation can be expressed as a stochastic Fourier integral with the assumption, that it is a steady random function of space. 
\begin{equation}
T(x) = \int_\lambda e^{i x_p k_p } dh(k)
\end{equation}
where $h(k)$ is a random function of $k_1, k_2, k_3$. The vectors ${\bf k}$ and ${\bf k'}$ denote two points in the wave-number space. The product of their increments at the same point is very small but non zero. It is defined by, 
\begin{equation}
<dh^{*}({\bf k}) dh({\bf k})> = \Phi({\bf k}) d{\bf k}
\end{equation} 
Here the asterisk indicates the complex conjugate. However, for the case of temperature which is a scalar fluctuation $\Phi({\bf k})$ depends only on $k$. The correlation function is then given by 
\begin{equation}
m(r) = <TT'> = 4 \pi \int_0^\infty k^2 {\Phi(k)} \frac{Sin k r}{k r} dr
\end{equation}
The power spectrum of temperature fluctuations is thus defined by, 
\begin{equation}
G(k) = 4 \pi k^2 \Phi(k)
\end{equation} 
The relation between the power spectrum and the temperature correlation is given by, 
\begin{equation}
G(k) = \frac{2}{\pi} \int_0^\infty m(r) k r {Sin k r} dr
\end{equation}
One can obtain the power spectrum of the temperature fluctuations as long as we know the temperature at different points. Unfortunately, in this case it has not been possible to obtain the power spectrum of the temperature fluctuations in the two different planes as we have done for the energy fluctuations. So we start of with the assumption that the temperature fluctuations are isotropic. However, as we see from the results, the fluctuation spectrum does not turn out to be a Gaussian as is expected for the isotropic case. We see that the temperature fluctuation cannot be fitted by a Gaussian and hence conclude that the temperature fluctuations cannot be isotropic. There is thus an anisotropy in the temperature fluctuation as well.  

To obtain the power spectrum, we have to obtain the energy of the system at different length scales. The temperature can be calculated using the equation, 
\begin{equation}
\epsilon(x,y) = 12 ( 4 + 3 N_{f}) (\frac{T^4}{\pi^2})
\end{equation}
$N_f = 3$ here, is the number of quark flavors. The energy is calculated from the partons after they undergo scattering and so the temperature is the temperature of the pre-equilibrium stage. We plot the power spectrum of the temperature fluctuation at $\sqrt{s} = 200$ GeV at two different times. 

\begin{figure}
\includegraphics[width = \linewidth] {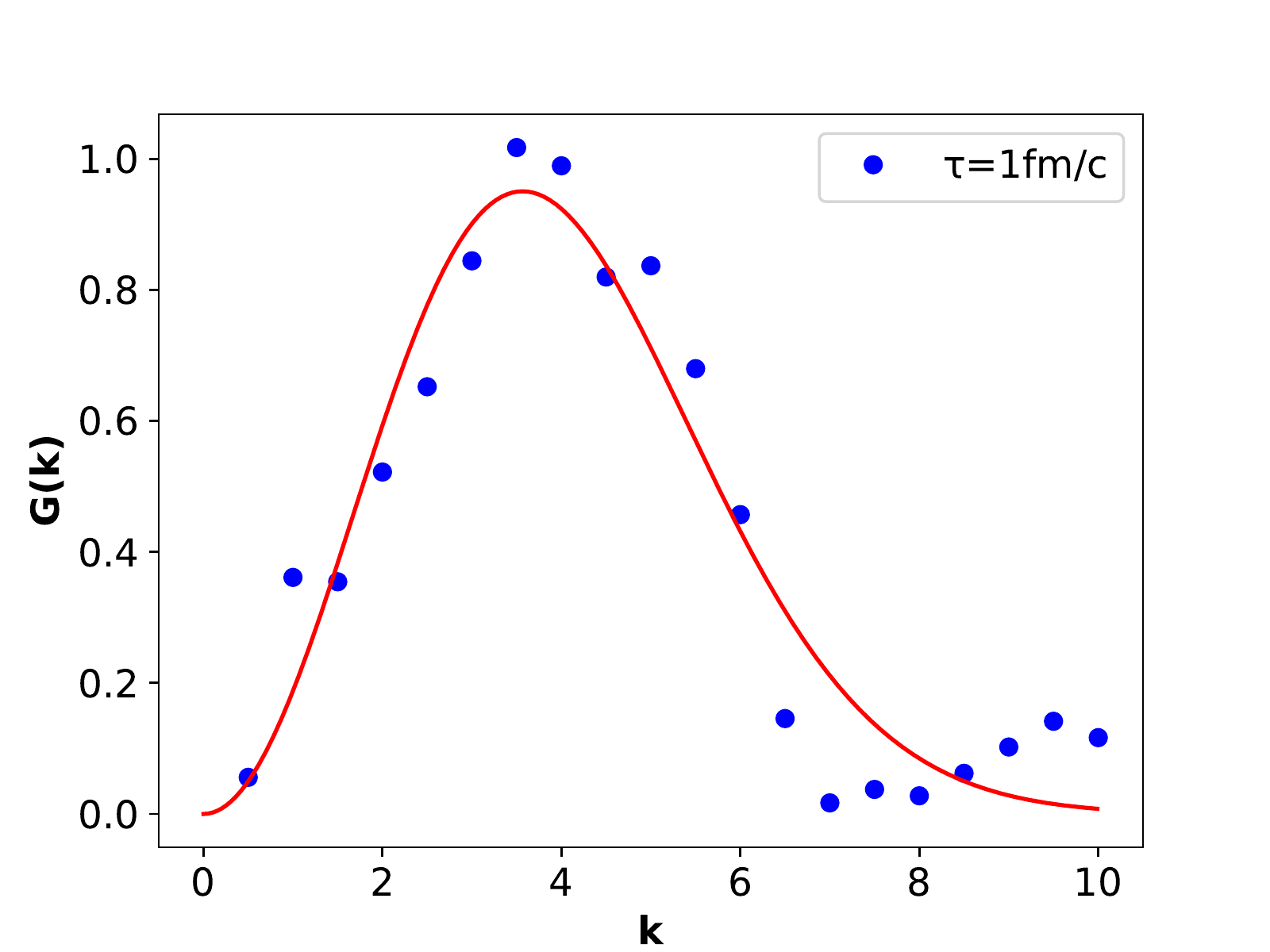}
\caption{The power spectrum of the temperature fluctuations for $200$ GeV Au-Au central collision events at $\tau = 1fm/c$. The units of $k$ is in $fm^{-1}$. It is fitted with a Gaussian function. }
\label{fig:t1}
\end{figure}
\begin{figure}
\includegraphics[width = \linewidth]{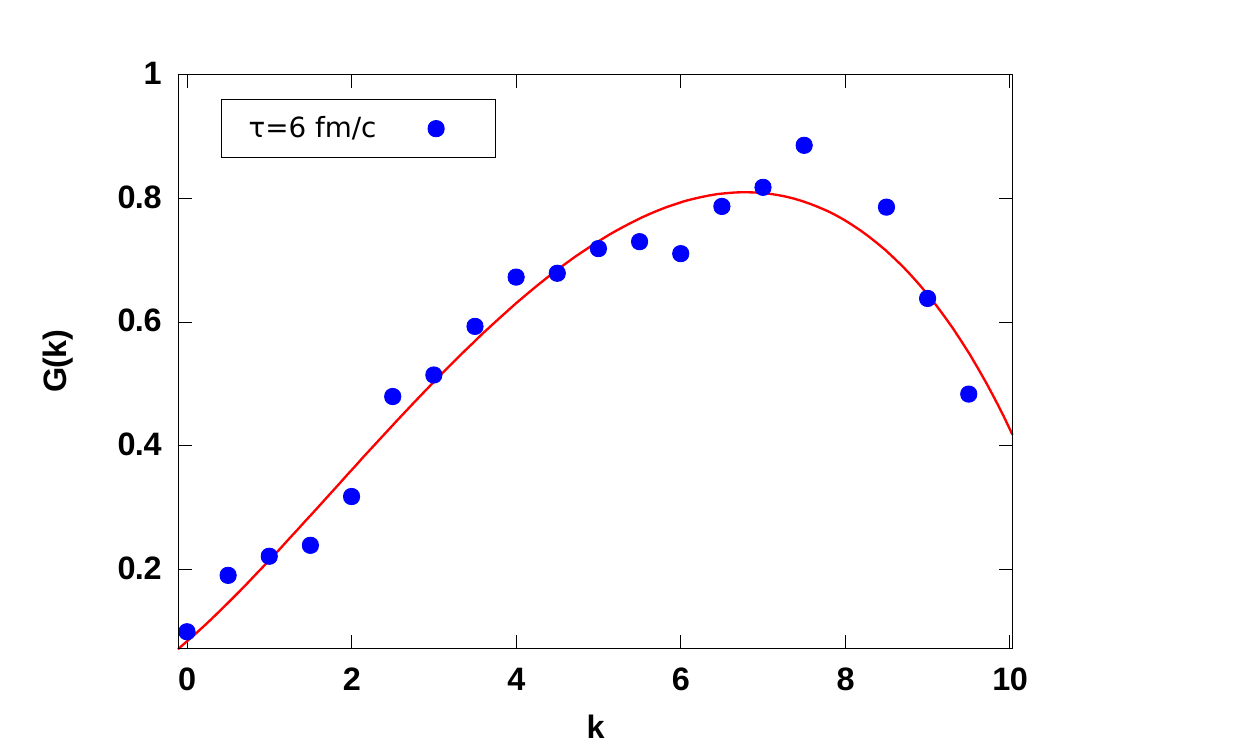}
\caption{The power spectrum of the temperature fluctuations for $200$ GeV Au-Au central collision events at $\tau = 6fm/c$. The units of $k$ is in $fm^{-1}$. At later times the peak of the power spectrum shifts to higher values of $k$, it is fitted with an asymmetric Poissonian q -Gaussian distribution. }
\label{fig:t6}
\end{figure}

Fig \ref{fig:t1} shows the power spectrum of the temperature fluctuations at $\tau = 1 fm/c$. This means that in the numerical simulations, we have taken all the parton scatterings upto $\tau = 1 fm/c$ to obtain the power spectrum.  We find that it 
can be approximately fitted with a Gaussian distribution. However, we do see that the fit is not so good at larger $k$ values which indicates smaller length scales. We find that the peak of the Gaussian shifts to higher $k$ values and therefore lower length scales as time progresses. 
Fig  \ref{fig:t6} shows the power spectrum of the temperature fluctuations at $\tau = 6 fm/c$.  Again this means that we have considered all the parton scatterings upto $\tau = 6 fm/c$ to obtain the spectrum of the temperature fluctuations.
Interestingly, we find that this nature persists at all the collision energies that we have studied. The shift to smaller length scales seems to indicate that energy is transferred to smaller eddies as time progresses. 
The scale of the temperature fluctuation can also be estimated, from the equation, 
\begin{equation}
\lambda_{sc} = \frac{1}{<T^2>_{avg}} \int_0^{\infty} m(r) dr
\end{equation} 
Our estimated scale of the temperature fluctuation is $1.16$ fm at $\tau = 6 fm/c$. This length scale is similar to the length scale calculated for the smallest eddies in the previous section.

Though the power spectrum of the temperature fluctuations can be approximately fitted by a  Gaussian even at later times, a better fit for the latter time points is the q-Gaussian
distribution. The q-Gaussian distribution is a generalization of the standard normal probability density. Since at latter times, the distribution becomes asymmetric, we use the asymmetric Poissonian q-Gaussian distribution \cite{pre} to fit the given distribution.
The fit to the distribution using an asymmetric q-Gaussian distribution is shown in Fig. \ref{fig:t6}. The distribution itself is defined by two parameters,  $q$ and the asymmetric parameter $a$. The distribution is given by, 
\begin{equation}
P(k) = \frac{c}{N_{\beta \beta'}}(1-k^2)^{((\beta + \beta')/2)-1}  \left(\frac{1+k}{1-k} \right)^{(\beta - \beta')/2}
\end{equation} 
Here we have, 
\begin{equation}
q = 1 - \left(\frac{(\beta + \beta')}{2}-1\right)^{-1}
\end{equation}
and the asymmetry parameter is given by, 
\begin{equation}
a = \frac{(\beta - \beta')}{2}
\end{equation}
Here $N_{\beta \beta'}$ is a constant given by, 
\begin{equation}
N_{\beta \beta'} = 2^{\beta +\beta'-1} \Gamma(\beta)\Gamma(\beta')/\Gamma(\beta+\beta').
\end{equation}
 
\noindent{In this case, we have fitted the temperature spectrum by a distribution based on three constants $c$, $\beta$ and $\beta'$}. 

It seems that though it is possible to fit an approximate Gaussian to the temperature fluctuations but at later times the anisotropy increases. For the temperature fluctuations, it was not possible to look at individual planes. However, the temperature spectrum of turbulence also indicates that the turbulence in the relativistic heavy-ion collisions is anisotropic.

\section{Conclusions}

In conclusion,  we have looked at the partonic system in the initial stages and the pre-equilibrium stages of the heavy-ion collision.  We have obtained the energy spectrum for the turbulent flow velocities. The range of the wave number and hence the eddy sizes are first obtained for the particular system. 
We have then obtained the turbulence spectra on the transverse plane of the heavy-ion collision as well as on the longitudinal plane at different initial conditions. We have obtained the coefficients of the spectra, $\nu$ for the two different planes. We have found that in the case of the longitudinal spectra, the $\nu$ value is $-5/3$ which is higher compared to the transverse plane spectra. For the transverse plane spectra, the $\nu$ value is coming out to approximately $-4/3$. The $\nu$ value of the longitudinal plane is therefore closer to the value of the Kolmogorov spectra. In the case of the transverse spectra, more energy dissipation takes place. We also found that the coefficient of the power law depends on the centrality of the collision only for the transverse plane. The reason for this could be the anisotropic pressure gradients generated in this plane due to the spatial anisotropy of the collision. The particles are also distributed anisotropically in this plane. So the viscosity has a centrality dependence which is reflected in the Schmidt number of the spectrum.   

The energy spectrum indicates that energy is transferred from the flow to the large eddies (low wave number) and is then dissipated through the smaller eddies in the heavy-ion collision system. Though there may be an overall isotropy in the turbulent system, if we slice the system into different planes, the coefficients of the power law vary between the different planes of the system. We show this variation between the transverse and the longitudinal plane. In most of the discussions on turbulence in relativistic heavy-ion system, the system is considered to be isotropic, however if we take the geometry of the collision into account the turbulence studied will always be anisotropic. This is well established in our study. We also discuss an anisotropy parameter $A$ which is based on the energy spectrum of the different planes of the fireball in the initial stages. The development of turbulence and the signals of turbulence should depend on this anisotropic parameter. We plan to pursue this development in a subsequent work.  

We also study the temperature spectrum of the turbulent quark gluon plasma in the pre-equilibrium stage. To our knowledge, this has not been studied before. The importance of the temperature spectrum lies in the fact that velocity correlations and temperature correlations need not be the same in a given system even though they might be related to each other. The temperature spectrum explores the thermal lengthscales in the underlying problem. In a high temperature plasma we feel that understanding of these thermal lengthscales are also quite important. We have found that though the thermal spectrum appears to be Gaussian but at later times it is better fitted with a q-Gaussian distribution. This indicates that though there is an overall isotropy and homogeneity to the turbulence observed in relativistic collisions, a better way of looking at this turbulence would be to slice it into planes and study the individual planes separately. This will help us to understand the anisotropies generated in the turbulent plasma on different planes and different lengthscales. We hope to pursue our understanding of the thermal spectrum in detail in a later work.

\begin{center}
 Acknowledgments
\end{center}  
 For computational infrastructure, we acknowledge the Center for Modeling, Simulation and Design (CMSD) at the University of Hyderabad, where part of the simulations was carried out. A.S is supported by INSPIRE Fellowship of the Department of Science and Technology (DST) Govt. of India, through Grant no: IF170627. The authors would like to thank the referees for their constructive comments which has improved the paper to a large extent.

\end{document}